# *Ultrahigh-Gain Phototransistors Based on Atomically Thin Graphene-MoS$_2$ Heterostructures*

A phototransistor based on the graphene/MoS$_2$ heterostructure is able to provide a high photoresponsivity greater than $10^7$ A/W.

*By Wenjing Zhang[†], Chih-Piao Chuu[†], Jing-Kai Huang[†$], Chang-Hsiao Chen[†], Meng-Lin Tsai[&], Yung-Huang Chang[†], Chi-Te Liang[#], Yu-Ze Chen[^], Yu-Lun Chueh[^], Jr-Hau He[&], Mei-Ying Chou[†#%\*] and Lain-Jong Li[†//\*]*

[†] *Institute of Atomic and Molecular Sciences, Academia Sinica, Taipei, 11529, Taiwan*

[$] *Department of Photonics, National Chiao Tung University, HsinChu 300, Taiwan*

[&] *Graduate Institute of Photonics and Optoelectronics, and Department of ElectricalEngineering, National Taiwan University, Taipei, Taiwan*

[#] *Department of Physics, National Taiwan University, Taipei, Taiwan*

[^] *Department of Materials Science and Engineering, National Tsing-Hua University, Hsinchu, 300, Taiwan*

[%] *School of Physics, Georgia Institute of Technology, Atlanta, GA 30332, USA*

[//] *Department of Physics, National Tsing Hua University, HsinChu 300, Taiwan*

\* To whom correspondence should be addressed: (L.J.Li) lanceli@gate.sinica.edu.tw; (M.Y. Chou) mychou6@gate.sinica.edu.tw

**Due to its high carrier mobility, broadband absorption, and fast response time, graphene is attractive for optoelectronics. However, the extraction of photoelectrons in conventional metal-graphene junction limits their photoresponsivity, typically lower than 0.01 AW$^{-1}$. Here we show that a large-area and continuous molybdenum disulfide (MoS$_2$) monolayer is achievable using a**

**CVD method and graphene is transferable onto MoS$_2$. We demonstrate that a phototransistor based on the graphene/MoS$_2$ heterostructure is able to provide a high photoresponsivity greater than $10^7$ A/W. Our experiments show that the electron-hole pairs are produced in the MoS$_2$ layer after light absorption and subsequently separated across the layers. Contradictory to the expectation based on the conventional built-in electric field model for metal-semiconductor contacts, photoelectrons are injected into the graphene layer rather than trapped in MoS$_2$ due to the presence of a perpendicular effective electric field caused by charged impurities or adsorbates, resulting in a tuneable photoresponsivity.**

Two-dimensional (2d) nanomaterials, such as graphene and MoS$_2$, hold great promise in next-generation electronic and photonic applications because of their unique properties inherited from the ultrathin planar structures, such as strong electron-hole confinement, extreme bendability, and high transparency, which allow for the fabrication of thinner, more flexible and more efficient devices[1,2]. Graphene has attracted substantial attention in optoelectronic applications due to its high carrier mobility, broad absorption spectrum, and fast response time. Graphene can absorb light and turn it into a photocurrent, and a recent study has shown that graphene serves as an excellent light-to-current converter with a quantum efficiency reaching close to 100% owing to its long mean-free path and high Fermi velocity[3]. However, graphene absorbs only 2.3% of light in the wide range of visible spectra[4]. Various approaches such as graphene plasmons[5,6], microcavities[7,8] and metallic plasmons[9] have been employed to enhance light absorption in graphene, but no obvious photogain has been reported for these graphene-based photodetectors. The photoresponse mechanisms in various types of graphene-based devices have been identified, including the photovoltaic effect[10-15], the thermoelectric Seebeck effect[15-19] and the bolometric effect[20]. The main reasons for the low photoresponsivity of graphene (~$1\times10^{-3}$ A/W) are the fast recombination of photoexcited carriers and the difficulty to create large enough p–n or metal-graphene junction areas, which are necessary for the electron–hole separation in 2d materials.

Assembling graphene with various 2d layers into artificial heterostructures to demonstrate new or tailored properties has been proposed[21] and realized in tunneling field-effect transistors[22-24] very recently. The photoresponse efficiency of graphene devices, in principle, can be greatly enhanced by exploiting a vertical geometry, for instance, graphene/2d semiconductor heterostructural stacking, where the whole graphene area can be used as a junction. The layered molybdenum disulfide ($MoS_2$) is a newly emerging 2d nanomaterial with a direct and finite band gap. Recent reports have demonstrated a gigantic photoluminescence (PL) from the $MoS_2$ monolayer, 4-fold higher than that in its bulk, owing to the quantum confinement effect associated with the transition from an indirect band gap in the bulk to a direct band gap in the monolayer[25-31]. The photodetectors based on $MoS_2$ thin layers have shown reasonably high photoresponsivity in ambient ranging from 7.5 mA/W to 780 A/W[32-34]. Very recently, the graphene/$WS_2$/graphene heterostructural device with a $WS_2$ thickness of 5-50 nm has also been demonstrated to exhibit a photoresponsivity ~ 0.1A/W.[35] In this work, we fabricate a phototransistor based on a graphene-on-$MoS_2$ heterostructure, where the $MoS_2$ monolayer grown by chemical vapor deposition (CVD) is used to absorb light and produce electron-hole pairs. The photo-excited electron-hole pairs are separated at the $MoS_2$ and graphene interfaces, where the electrons move to graphene due to the presence of a perpendicular effective electric field created by charged impurities or adsorbates. The phototransistor based on the graphene/$MoS_2$ heterostructure is able to reach a photoresponsivity value higher than $10^7$ A/W and a photogain of about $10^8$.

In our previous work[36], we have reported the direct growth of $MoS_2$ monolayer crystal flakes on a sapphire or $SiO_2$ substrate by the vapour-phase reaction of $MoO_3$ and S powders in a hot-wall chemical vapour deposition (CVD) system. Here we report that this method can be further extended to grow on $SiO_2$ a continuous $MoS_2$ layer composed of randomly oriented crystalline $MoS_2$ domains with an average domain size around several microns. The $MoS_2$ films were synthesized on cleaned sapphire substrates in a hot-wall furnace. High purity $MoO_3$ (0.3g; from Aldrich; 99% purity) was placed in a ceramic boat at the heating center of the furnace). Sapphire Substrates were placed beside the ceramic boat as shown in the supporting Figure S1. Sulfur powder was heated by heating tape (160°C) and carried by Ar (Ar =70 sccm at 10 torr) to the furnace heating center. The furnace was gradually heated from room temperature to 650°C with a rate of 25°C/min. After keeping at 250°C for 10 minutes, the furnace was naturally cooled down to room temperature. For the transfer of the $MoS_2$ layer, $MoS_2$/sapphire was coated with a layer of PMMA (Micro Chem. 950K A4) by spin-coating (step1: 500 rpm for 10 sec; step 2: 3000 rpm for 60sec), followed by a baking at 100°C for 10 min.  The PMMA-supported $MoS_2$ was dipped into a NaOH (2M) solution at 100°C for 30min. It was then detached and transferred to the de-ionized water for the removal of the NaOH. A fresh $SiO_2$/Si substrate was then used to fish the PMMA-supported $MoS_2$ film, followed by drying on a hot-plate at 100°C for 10 min.  The PMMA was removed by acetone and isopropyl alcohol. The as-grown $MoS_2$ was mostly monolayer although we also noticed that the growth of small-sized second layer of $MoS_2$ was initiated at the center of some monolayer domains. Large-area monolayer graphene was grown on copper foils at 1000 °C by a CVD method using a mixture of methane and hydrogen gases as reported elsewhere[37,38]. To stack the graphene monolayer on MoS$_2$, a layer of PMMA thin film was coated on the graphene/Cu foil as a transfer supporting layer[39,40]. After the wet etching of Cu by an aqueous solution containing Fe$^{3+}$ ions, the PMMA-supported graphene film was transferred to the top of the as-grown MoS$_2$ film on SiO$_2$/Si, followed by the removal of PMMA.

The photograph in Figure 1a is the top view of the graphene/MoS$_2$ heterostructure simply formed by a manual stacking of a large-area CVD graphene monolayer onto MoS$_2$. In Figure 1a, monolayer MoS$_2$ film was directly grown on the right hand side of the SiO$_2$/Si wafer followed by the transfer of graphene to the bottom half; therefore, we can see a clear difference in the optical contrast at four quadrants. Figure 1b schematically illustrates the device structure adopted in the study, where the top view of the comb-shaped source and drain metals is also shown below. We have performed the TEM cross-section study for the graphene/MoS$_2$ device. Figure 1c shows the TEM cross-section view, where the graphene/MoS$_2$ is capped with a passivation layer of SiO$_2$, and the intensity profile on the right hand side shows that the thicknesses of graphene and MoS$_2$ are as expected, 0.36 nm and 0.7 nm. There is no PMMA residue in between graphene and MoS$_2$ layers. Figure 1d displays the Raman spectrum of the MoS$_2$ monolayer on SiO$_2$/Si and that of MoS$_2$ covered by graphene, taken from the sample shown in Figure 1a. The energy difference between the Raman E$^1_{2g}$ and A$_{1g}$ peaks is ~ 19.0 cm$^{-1}$, indicating that the MoS$_2$ film is monolayer[26-28]. The optical micrograph of a MoS$_2$ film on a SiO$_2$ substrate is shown in supporting Figure S2a. The Raman mapping in supporting Figure S2b shows that the distribution of the energy difference between the Raman E$^1_{2g}$ and A$_{1g}$ peaks is uniform across the sample. The AFM cross-sectional height shown in supporting Figures S2c and S2d confirms that the film is monolayer. The peaks at about 2695.9 cm$^{-1}$ and 1581.5 cm$^{-1}$ are the characteristics of 2D and G bands, respectively for monolayer graphene[37]. Figure 1e illustrates that the photoluminescence (PL) spectrum for $MoS_2$ covered by graphene (graphene/$MoS_2$) maintains a similar shape as its pristine form (without graphene on top) except that the intensity is decreased.

Before discussing the photocurrent behavior, we examine the carrier properties when graphene contacts with $MoS_2$ in dark. First, we obtain the carrier concentrations and resistances of the graphene and graphene/$MoS_2$ sheets on $SiO_2$ by the Hall-effect measurements in dark. All the samples are the same size (0.5 cm $\times$ 0.5 cm), and each sample has been measured four times. Since $MoS_2$ is much less conductive compared with graphene, the carrier properties obtained are mainly from graphene. Note that significant numbers of reports have shown that CVD graphene is p-doped in ambient caused by the doping effect from transfer process, adsorbed moisture/oxygen and the substrate impurities [41-42]. Supporting Figures S3a and S3b demonstrate that the hole concentration in graphene decreases from $6 \times 10^{12}$ cm$^{-2}$ to $2 \times 10^{12}$ cm$^{-2}$ and the resistance significantly increases when graphene is in contact with $MoS_2$, suggesting that electrons possibly move from $MoS_2$ to graphene. Second, we prepared the graphene and graphene/$MoS_2$ transistors on a $SiO_2$/Si substrate with the same device fabrication processes. The electrical transfer curves in supporting Figure S3c demonstrate that the charge neutral point ($V_{CNP}$) of the graphene/$MoS_2$ transistor shifted to the left compared with that of a graphene transistor, indicating that the graphene/$MoS_2$ transistor was less hole-doped, consistent with the conclusion from Hall-effect measurements.

The Raman spectra in Figure 2a show that the G band of monolayer graphene on $SiO_2$/Si is at 1585.4 cm$^{-1}$ and it is broadened with a downshift to 1581.7 cm$^{-1}$ when a $MoS_2$ layer is present underneath. Note that the samples are continuously illuminated by a Raman laser during the measurement. Since the graphene transferred by PMMA is known as *p*-doped, the red shift and broadening of the G band indicate that the Fermi level of graphene is raised (or an increase in the electron concentration) with light exposure[43]. These results suggest that the photoelectrons generated by the Raman laser are injected into graphene. The Raman mappings of the G band energy and the FWHM are also shown to consolidate the conclusion. To further reveal the effect of adding graphene on MoS$_2$, Raman features for MoS$_2$ with and without graphene coverage are examined. The Raman spectra and mappings in Figure 2b show that the A$_{1g}$ peak is up-shifted in energy and the peak width is narrowed after being covered by graphene, indicating that the MoS$_2$ layer becomes less *n*-doped (or a decrease in the electron concentration)[44]. For comparison, we have also fabricated a stacking structure MoS$_2$/graphene by transferring MoS$_2$ monolayer onto a *p*-typed graphene layer on SiO$_2$ substrates. The Raman spectra and mappings shown in supporting Figure S4 demonstrate that when the graphene/MoS$_2$ stack is exposed to light, the graphene layer also receives electrons from MoS$_2$. Both structures consistently show that the photo-excited electrons move from MoS$_2$ to graphene, and the photo-excited holes are trapped in the MoS$_2$ layer.

To quantify the photocarriers, we study the dependence of the photocurrent on light power for the graphene/MoS$_2$ transistor. Figure 3a shows the transfer curves for the graphene/MoS$_2$ transistor exposed to the 650-nm light with various power densities (device structure shown in Figure 1b). The voltage of the charge neutral point V$_{CNP}$ for the transfer curve in dark is at around 10 V, indicating that graphene is *p*-doped. The shape of the transfer curve is very similar to that for pristine graphene on SiO$_2$, suggesting that the carrier transport in the graphene/MoS$_2$ phototransistor is dominated by graphene. The result is reasonable because graphene is much more conductive than the $MoS_2$ layer. The graphene/$MoS_2$ transistor is extremely sensitive to light. When light is illuminated on the graphene/$MoS_2$ transistor, the drain current ($I_d$) in the *p*-channel decreases and the $I_d$ in the *n*-channel increases as shown in Figure 3a, indicating that the photoexcited electrons are injected into graphene. At the same time, the $V_{CNP}$ largely shifted to a more negative voltage even with very weak light exposure (Figure 3b), where the photocurrent dependence on gate voltage is plotted in Figure 3c. The negative shift of $V_{CNP}$ indicates that the photoexcited holes were trapped in the $MoS_2$, acting as an additional positive gate voltage for graphene[45]. The carrier concentration of the trapped holes can be extracted using the formula $\Delta n = C_g \times \Delta V_{CNP}/e$ for graphene, where $C_g = 1.15 \times 10^{-8}$ F/cm$^2$ for the dielectric film of 300 nm $SiO_2$, and e is the electron charge[43,45,46]. The internal quantum efficiency (IQE), a measurement of a phototransistor's electrical sensitivity to light, can be estimated by the equation: IQE = (the number of photoexcited electron-hole pairs)/(absorbed number of photons) = $\Delta n \times$ A / ($P_o/h\nu$), where A, $P_o$, $h$ and $\nu$ represent the total channel area, absorbed light power by graphene and $MoS_2$, the Planck constant and the frequency of the incident laser, respectively. (Note that the absorbance of the graphene/$MoS_2$ heterostructure is ~ 6.8% at 650 nm based on our absorption measurement.) As shown in Figure 3d, the IQE decreases with the increasing light power and the largest IQE for the system is ~15%. We further estimate the photoresponsivity and photogain to quantify the photo sensitivity of the graphene/$MoS_2$ phototransistor. The photoresponsivity is the ratio between the photocurrent ($I_{ph} = I_{light} - I_{dark}$) and the light power absorbed by the phototransistor, and, remarkably, the photoresponsivity can reach $1.2 \times 10^7$ A/W (at $V_g$ = -10V; $V_{ds}$=1V; light power density ~0.01W/m$^2$) as shown in Figure 3e. The photogain can be calculated by the formula G = $I_{ph}$ / [e × (the number of photoexcited electron-hole pairs)] = $I_{ph}$ / (e × $\Delta n$ × A), and the gain is up to $10^8$ (Figure 3f). It is noteworthy to point out that the reported photoresponsivity for graphene and a pristine $MoS_2$ layer is around $10^{-3}$ and 780A/W [3,34], respectively. In the present experiment, the graphene-based phototransistor with an ultrahigh photo sensitivity is realized simply by stacking it onto an atomically thin $MoS_2$ layer.

In addition to the move of photoelectrons to graphene, the negative shift of $V_{CNP}$ upon photoexcitation may also be attributed to other extrinsic effects such as thermal desorption of absorbed dopants. Figure 4a shows the dependence of photoresponsivity of the graphene/$MoS_2$ transistor on the wavelength of light. It is observed that photoresponsivity only becomes pronounced when the excitation energy is higher than the absorption band gap of $MoS_2$ (1.8eV)[30], with the optical absorption feature of the as-grown $MoS_2$ layer shown in Figure 4b. These results suggest that the photocurrent is originated from the light absorption in $MoS_2$: the electron-hole pairs are produced in $MoS_2$, followed by the separation of them between $MoS_2$ and graphene layers. If the photocurrent were from the thermal effect, the photocurrent would have been induced even when the photon energy is smaller than the band gap of monolayer $MoS_2$.

To further exclude the thermal effect, we measured the time-resolved photocurrent driven by different drain voltages at a low laser power density, which ensures that the fast current self-heating effect becomes more prominent for the thermal desorption process. Figure 4c shows the normalized photocurrent-time profiles for the graphene/$MoS_2$ transistor in ambient air. It is observed that the current-time curve obtained at $V_{ds}$ = 0.002 V overlaps with that at $V_{ds}$ = 0.01V. The photocurrent becomes smaller when $V_{ds}$ is set at 1V (current $I_{ds}$ is ~2×$10^{-2}$A ). If the thermal desorption of dopants from the film were a dominant process, a higher $V_{ds}$ would have given rise to a larger photocurrent.[47] The measurement results in vacuum also consistently lead to the same conclusion that thermal desorption is not a dominant process. We have presented detailed results of time-resolved photoresponse and arguments in supporting Figure S5 to show that the photocurrent of the graphene/MoS$_2$ phototransistor is not from the thermal effects.

It is also crucial to examine whether the CVD graphene itself contributes to the observed large photocurrent. Our measured results in Figure 4d (and in supporting Figure S6) are consistent with those reposrted by Avouris' group [15] in that the intrinsic photoresponse in biased graphene transistors was dominated by the photovoltaic effect and a photoinduced bolometric effect, while the thermoelectric effect was insignificant. However, the photocurrent generated from the pure graphene transistors is several orders of magnitude smaller than that observed in our graphene/MoS$_2$ phototransistor, indicating that the observed large photocurrent from graphene/MoS$_2$ is not contributed by the aforementioned processes in graphene. After monolayer graphene and monolayer MoS$_2$ contact with each other, the electrons are injected into graphene based on our experimental results. Thus, the conventional theory of ideal metal-semiconductor contacts predicts an energy-band bending at the interface as shown in supporting Figure S7, and the direction of the built-in electric field is from MoS$_2$ to graphene. With illumination on the graphene/MoS$_2$ transistors, the photo-generated electrons should flow into MoS$_2$. However, this is opposite to our experimental results. Different from the conventional metal-semiconductor contact, the semimetal and semiconductor of the 2d heterostructure are atomically thin layers, and the depletion region for the bulk contacts does not exist. Thus, the interface is prompted to be modulated by an effective electric field mainly created by the charged impurities between graphene/MoS$_2$ or between MoS$_2$/substrates. When pumping the sample to high vacuum (~5×10$^{-5}$mbarr) in dark as presented in Figure 5a, the V$_{CNP}$ of the second device shifted to the left by ~23V, indicating that the adsorbates and impurities close to the MoS$_2$ layer are negative charged. So the direction of the effective electric field is from graphene to MoS$_2$, and with its help the photogenerated electrons flow into the graphene while the holes are trapped in the MoS$_2$ layer. Very recently, our research has shown that the adsorbates in ambient air can also *p*-dope the monolayer MoS$_2$ film[34]. However, the photogenerated electrons are injected into graphene for both graphene/MoS$_2$ and MoS$_2$/graphene structures, suggesting that the negative charge impurities are mainly associated with the MoS$_2$ layer.

As shown in Figure 5, we can find that the photocurrent in vacuum is much smaller than that in ambient air, and the photocurrent obtained at V$_g$-V$_{CNP}$ < 0 is consistently higher than that at V$_g$-V$_{CNP}$ > 0. It is expected that when the graphene/MoS$_2$ bilayer is created, the Fermi levels of the two layers should be aligned due to interlayer coupling, and energy band bending at the interface can be modulated by the doping level of the graphene and MoS$_2$ layers. To exploit this, first-principles calculations are performed based on density functional theory (DFT)[48,49] using the Vienna *ab initio* simulation package (VASP) [50,51]. The details are described in Methods. A modulated electric field ($E_{ext}$) is applied to simulate the doping effect and align the energy bands in graphene/MoS$_2$. The band structure corresponding to the bilayer in air is shown in Figure 6a with *p*-doped graphene and *n*-doped MoS$_2$. After pumping that removes *p*-dopants, the band structure of the bilayer in vacuum is shown in Figure 6b with *n*-doped graphene and *n*-doped MoS$_2$. The corresponding schematic illustration of the photoelectron transfer process is also shown in Figure 6. From the Figure 6, we can find that the effective electric field of the p-doped graphene/$MoS_2$ heterostructure would be larger than that of the n-doped graphene, so the photogenerated electron-hole pairs would be separated more easily by this effective electric field. The more *n*-doped graphene, the smaller external electric field, and leading to the smaller photogain. So the photoresponsivity of the graphene/$MoS_2$ phototransistor can be tuned by the modulation of the doping level of graphene and $MoS_2$. As the monolayer $MoS_2$ has a band gap ~1.8eV, and the graphene is semimetal, the photoresponsivity is more sensitive to the graphene doping level.

The high photogain process in the graphene/$MoS_2$ bilayer can now be described as follows. Light absorption in $MoS_2$ generates electron-hole pairs; the electrons can move to the graphene layer due to an effective electric field created by charged impurities or adsorbates, while the holes are trapped in the $MoS_2$ layer. The high electron mobility in graphene and the long charge-trapping lifetime of the holes result in multiple recirculation of electrons in graphene, leading to a very high photogain. This high-photogain mechanism is similar to what was reported by Konstantatos et al. for bilayer graphene where a thick layer of PbS quantum dots was used as the light absorber[45], although the controlling scheme for charge separation there is intrinsically different from that in the current heterostructure formed by 2d layered materials.

**CONCLUSIONS**

In conclusion, we have constructed a graphene/$MoS_2$ bilayer by manually stacking graphene on a CVD $MoS_2$ layer. The advantage of using this structure for photodetection is that the whole surface area can be used as a junction, where electron–hole pairs can be separated at the interface. The phototransistor based on this graphene/$MoS_2$ heterostructure is able to reach a photoresponsivity value higher than $10^7$ A/W while maintaining its unique ultrathin character. Since the device is very sensitive to weak light, it may be used as an ultrasensitive, ultrathin, and flexible photodetector.  Our results suggest that the consideration of interlayer coupling leading to the alignment of the Fermi level in two layers better explains the photocurrent behaviour than the conventional built-in electric field (photovoltaic) model.  The present work demonstrates the significance of charge movement in the emerging field of 2d heterostructures. The heterostructures of 2d layers exhibit novel materials properties beyond the capacities of the constituents. Stimulations of research and developments of optoelectronic applications based on various heterostructural 2d materials are thus anticipated.

**Acknowledgements:**

This research was supported by Academia Sinica (IAMS and Nano program) and National Science Council Taiwan (NSC-99-2112-M-001-021-MY3).

**Figures**

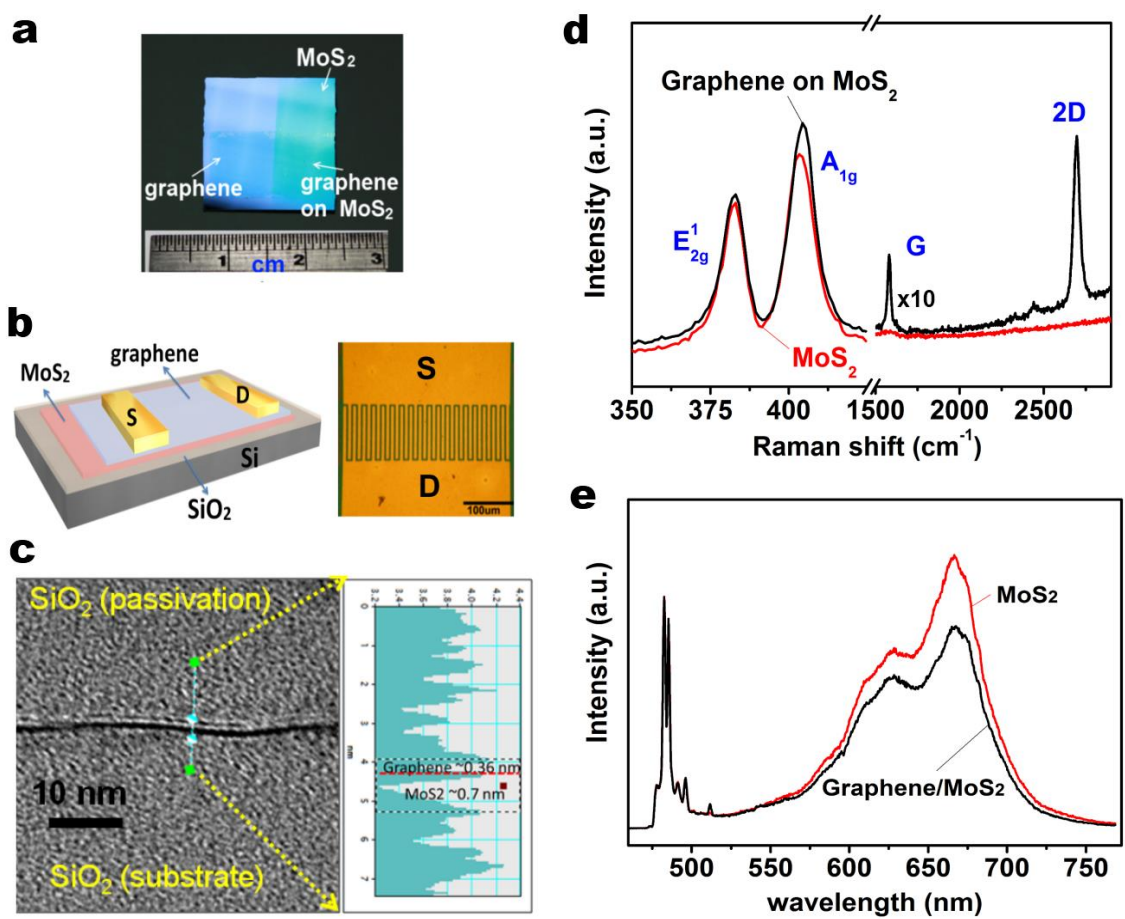

**Figure 1. Spectroscopic characterizations of a graphene/MoS$_2$ bilayer.** (a) Photo showing that a MoS$_2$ monolayer was grown on the right hand side of the 300 nm SiO$_2$/Si wafer followed by transferring graphene onto the bottom half. (b) Schematic illustration of the phototransistor based on graphene/MoS$_2$ stacked layers, where the channel is formed in between the comb-shaped source and drain metal electrodes (Ti/Au = 5nm/80nm). (c) A high-resolution TEM (HRTEM) image clearly reveals a bilayer stacking of graphene/MoS$_2$. The thickness of each layer can be extracted from the intensity profile to be 0.36 nm and 0.7 nm for graphene and MoS$_2$, respectively. (d) Raman spectra and (e) photoluminescence spectra for MoS$_2$ and MoS$_2$ covered by CVD monolayer graphene taken from the sample shown in (a). Note that the Raman intensity of Graphene/MoS$_2$ in (d) has been multiplied by a factor of 10 for better comparison.

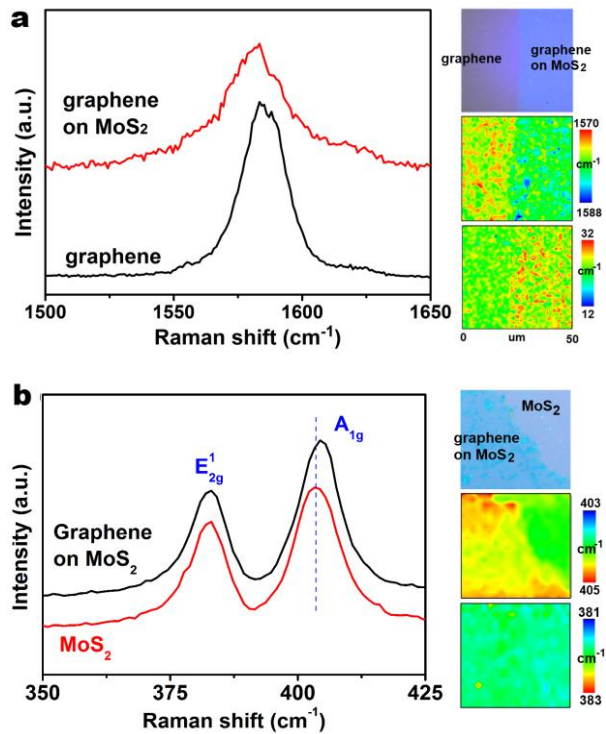

**Figure 2. Raman characterizations.** (a) Raman mappings and representative Raman spectra for graphene and graphene on MoS$_2$, where the G band energy of graphene on SiO$_2$ is higher than that transferred on MoS$_2$. (b) Raman mappings and representative Raman spectra for MoS$_2$ and that covered with graphene. The A$_{1g}$ energy of MoS$_2$ is up-shifted when graphene is transferred onto it, suggesting that the electron density in MoS$_2$ is lowered (with electrons moving from MoS$_2$ to graphene). The excitation source is a 473-nm laser.

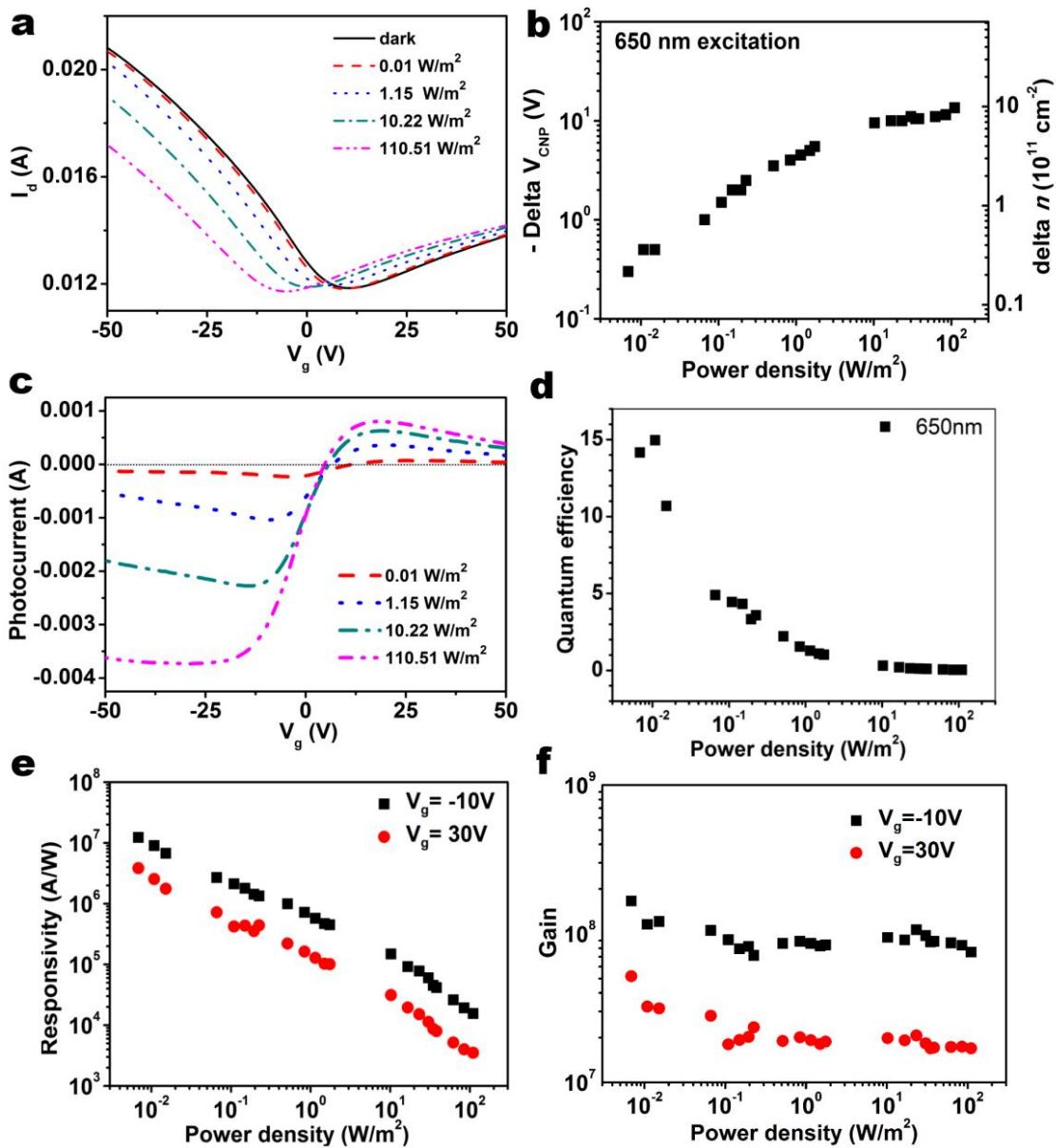

**Figure 3. Photoresponses of the graphene/MoS$_2$ devices.** (a) Transfer curves for the graphene/MoS$_2$ phototransistors under the exposure of light with various powers. (b) Shift of the charge neutral point V$_{CNP}$ and the corresponding electron density change (delta n) for a graphene/MoS$_2$ phototransistor with various light powers. (c) Photocurrent as a function of the gate voltage based on the transfer curves obtained in (a). (d) Quantum efficiency, (e) photoresponsivity and (f) photogain for the graphene/MoS$_2$ phototransistors. The wavelength of the laser is 650 nm, and the channel area for exposure is ~$2.0 \times 10^{-8}$ m$^2$.

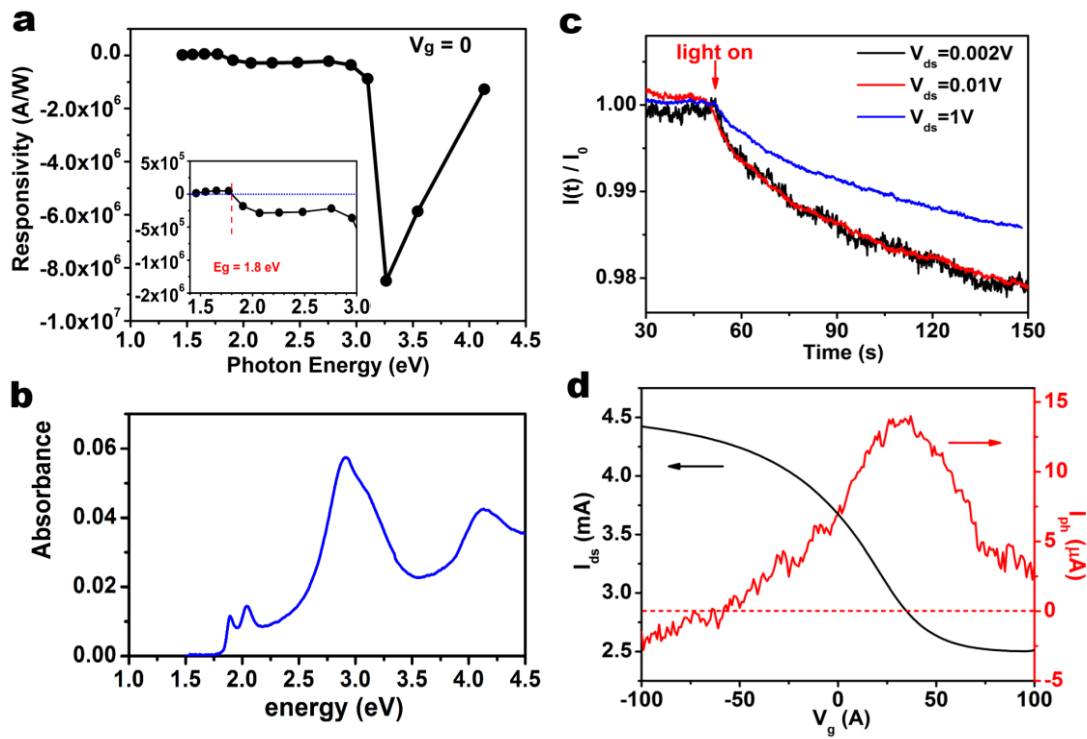

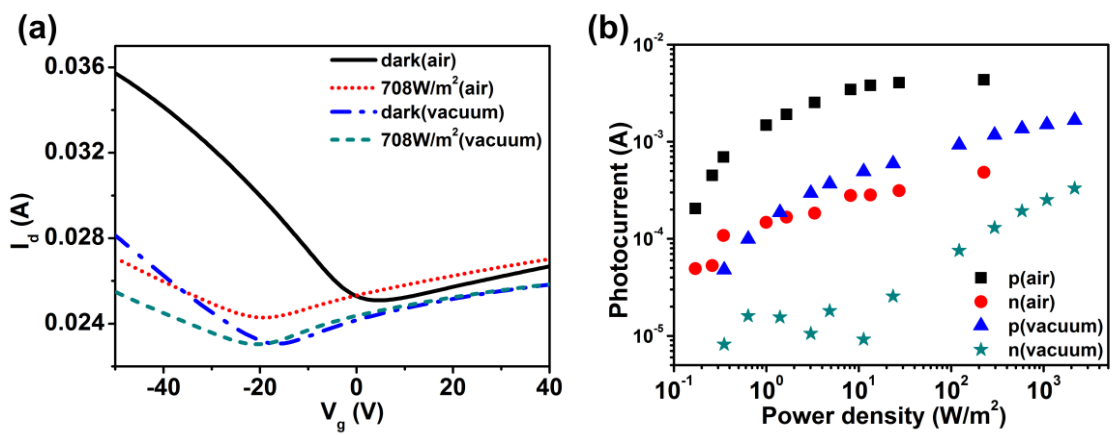

**Figure 4. Study of photresponse mechanisms.** (a) Photoresponsivity as a function of the energy of the excitation light source. (b) Optical absorption spectrum for monolayer $MoS_2$. (c) The normalized photocurrent-time profiles the graphene/$MoS_2$ transistor measured with various drain voltages in ambient air, where current $I(t)$ is the measured current divided by the current $I_0$ when light is turned on. A continuous wavelength 532 nm laser was used to illuminate the device at power density 0.34 W/m$^2$. (d) The photocurrent of a phototransistor based on a CVD graphene layer. A continuous wavelength 532 nm laser was used to illuminate the device at power density 35.21W/m$^2$ ($V_{ds}$=0.1V).

**Figure 5. Comparison of photoresponses in air and in vacuum.** (a) Transfer curves for the graphene/MoS$_2$ transistor measured in air and in vacuum. (b) Photocurrent as a function of the light power density in air and in vacuum at $V_g$-$V_{CNP}$=±20V. The 532-nm laser was used to measure the photocurrent, and the spot size was ~2 mm.

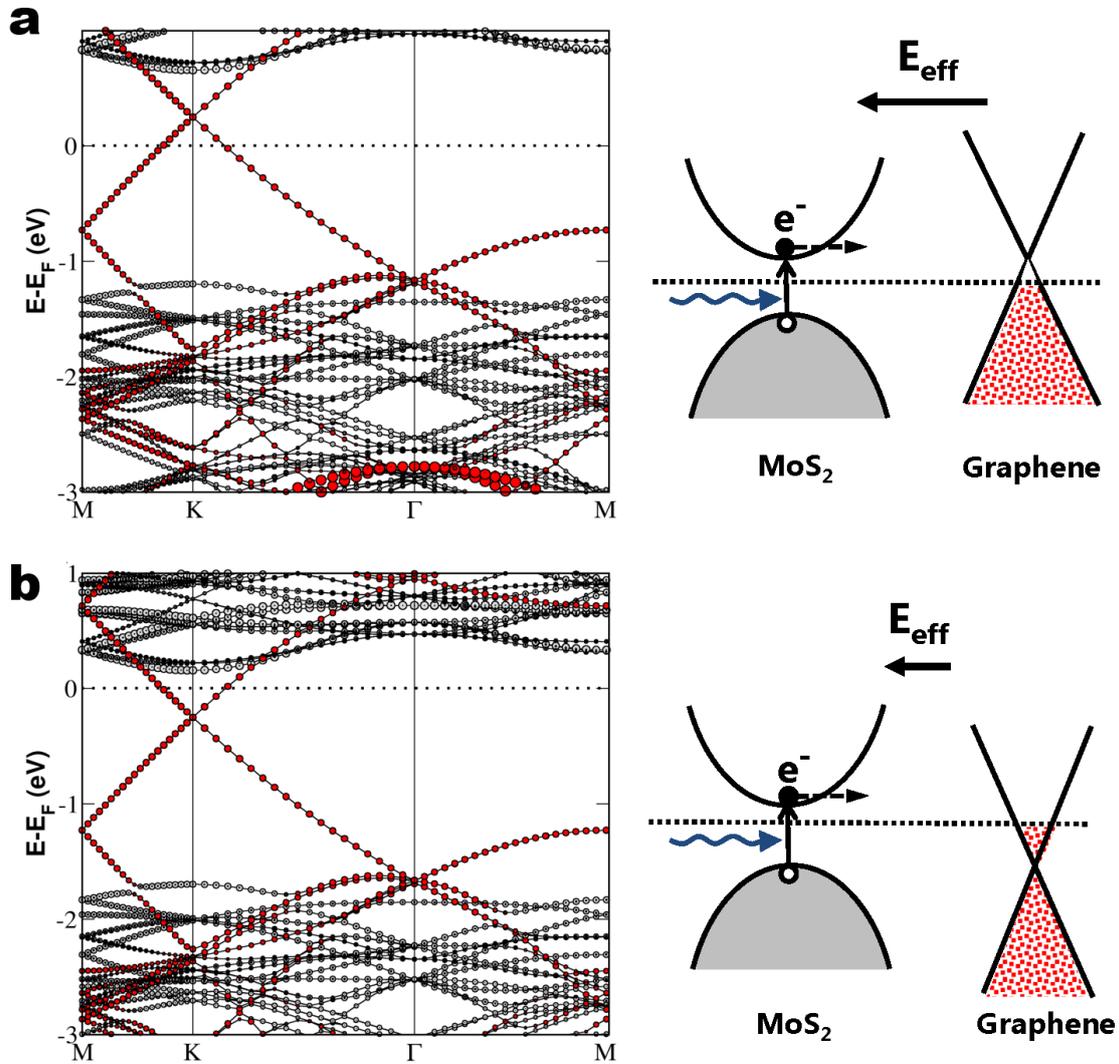

**Figure 6. Schematic illustration of the photoelectron transfer process in the graphene/MoS$_2$ bilayer.** The corresponding band structures for an *n*-doped MoS$_2$ layer topped with (a) slightly *p*-doped graphene and (b) *n*-doped graphene corresponding to the experimental situation in air and in vacuum, respectively. The electronic states associated with graphene and MoS$_2$ are represented in red and gray in the band structure plots, respectively. The effective electric field is created by charged impurities or adsorbated (see text).

# Supporting Materials

## Ultrahigh-Gain Phototransistors Based on Atomically Thin Graphene-MoS$_2$ Heterostructures

By Wenjing Zhang[†], Chih-Piao Chuu[†], Jing-Kai Huang[†$], Chang-Hsiao Chen[†], Meng-Lin Tsai[&], Yung-Huang Chang[†], Chi-Te Liang[#], Yu-Ze Chen[^], Yu-Lun Chueh[^], Jr-Hau He[&], Mei-Ying Chou[†#%*] and Lain-Jong Li[†//*]

[†] Institute of Atomic and Molecular Sciences, Academia Sinica, Taipei, 11529, Taiwan

[$] Department of Photonics, National Chiao Tung University, HsinChu 300, Taiwan

[&] Graduate Institute of Photonics and Optoelectronics, and Department of ElectricalEngineering, National Taiwan University, Taipei, Taiwan

[#] Department of Physics, National Taiwan University, Taipei, Taiwan

[^] Department of Materials Science and Engineering, National Tsing-Hua University, Hsinchu, 300, Taiwan

[%] School of Physics, Georgia Institute of Technology, Atlanta, GA 30332, USA

[//] Department of Physics, National Tsing Hua University, HsinChu 300, Taiwan

* To whom correspondence should be addressed: (L.J.Li) lanceli@gate.sinica.edu.tw; (M.Y. Chou) mychou6@gate.sinica.edu.tw

# 1. Methods

**Characterizations**: The AFM images were performed in a Veeco Dimension-Icon system. Raman and photoluminescence (PL) spectra were collected in a confocal Raman/PL system (NT-MDT). The wavelength of laser is 473 nm (2.63eV), and the spot size of the laser beam is ~0.5μm. The step size of the Raman spatial mapping is 0.5 $\mu$m, and the spectral resolution is 3 cm$^{-1}$ (obtained with a 600 grooves/mm grating). A high grating (1800 grooves/mm) is also used to get more details of the line shapes of the Raman band, and the spectral resolution is 1cm$^{-1}$. The Si peak at 520 cm$^{-1}$ was used as a reference for wavenumber calibration, and the peak frequency was extracted by fitting a Raman peak with a Lorentz function. The electrical measurements were performed in an ambient or vacuum condition using a Keithley semiconductor parameter analyzer, model 4200-SCS. The sheet resistance, sheet concentration, and mobility of the graphene and graphene on MoS$_2$ films were analyzed by using a Hall sensor measurement based on the Van der Pauw method[R1]. Field-emission transmission electron microscopy (JEM-3000F, JEOL operated at 300 kV with point-to-point resolution of 0.17 nm) equipped with an energy dispersion spectrometer (EDS) was used to obtain the microstructures and the chemical compositions. The sample for TEM measurements was prepared by a focus ion beam (FIB) (SMI 3050SE, SII Nanotechnology). Before sample cutting, the sample was capped by a 100 nm-thick SiO$_2$ layer deposited by an E-gun system as a passivation layer in order to prevent the damage from Ga ions during sample cutting.

**Photocurrent measurement.** The following figure shows the schematic of our photoresponse measurement setup. The spot size of the 532-nm and 650-nm lasers is about 2 mm and 1mm, respectively. The strongest laser power intensity was about 8.8 mW (power density: 2.8×10$^3$ W/m$^2$), and the weakest was about 30 nW (power density: 1×10$^{-2}$ W/m$^2$). The neutral density filters were used to adjust the light power. To prevent interference from miniscule room light, we used a lightproof hood to cover up the system, and the measured power density of background light was about 3×10$^{-4}$ W/m$^2$; thus, the interference from the background light can be neglected. To reduce the interference from the reflected light of the laser, the diameter of the quartz window was set to be around 5 mm. We also measured the background light in the vacuum chamber when the laser was on and found that it was smaller than 1×10$^{-3}$ W/m$^2$, still one order of magnitude smaller than the lowest power density used in the experiments. The spectral responsivity of the graphene/MoS$_2$ phototransistor was measured with the EQE-R3001 spectral response system (Enli Technology Co., Ltd.) under V$_{ds}$=1V at room temperature in ambient air.

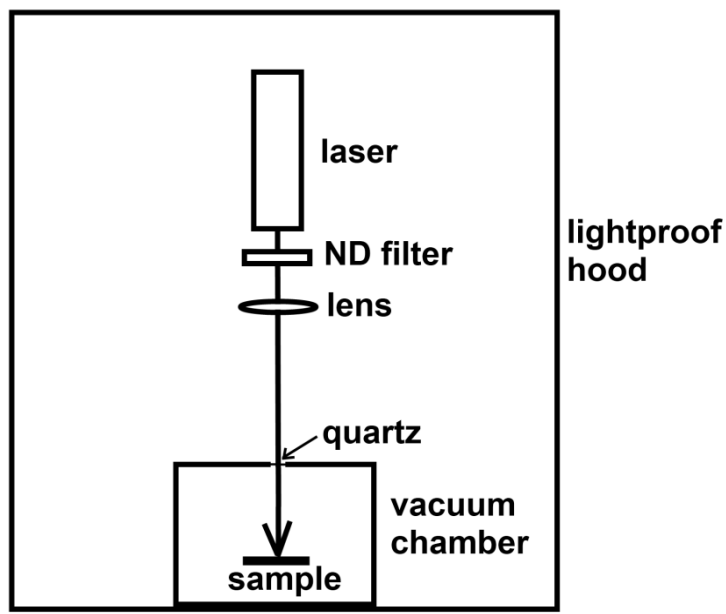

**Schematic of the photoresponse measurement setup.**

**Numerical Simulation:** We construct a slab model with a 5×5 supercell of graphene on a 4×4 supercell of MoS$_2$ to simulate the stacked layer, where the interactions between electrons and ions is described by the projector augmented wave (PAW)[R2] method, and the exchange-correlation potential is described by the local density approximation (LDA)[R3]. The lattice constant of graphene used in the simulation is 2.46 Å. The lattice constant for optimized monolayer MoS$_2$ is 3.12 Å. A vacuum thickness of 15 Å is used in order to eliminate the spurious image interactions, and the energy cut-off of plane waves is 400 eV. The interlayer spacing between MoS$_2$ and graphene is 3.3 Å for the optimized structure. An applied electric filed is added to simulate the effects of an applied gate voltage or Coulomb impurities. The calculated band gap of MoS$_2$ is 1.8 eV, which is known to be underestimated by LDA.

**Growth MoS$_2$ monolayers.**

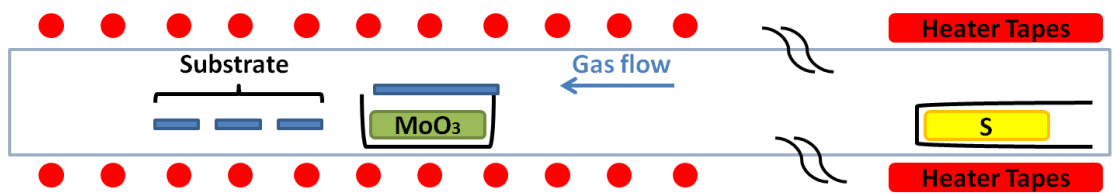

**Figure S1. Schematic illustration of the furnace set-up for the MoS$_2$ growth.**

**Characterization of monolayer MoS$_2$ grown by CVD.**

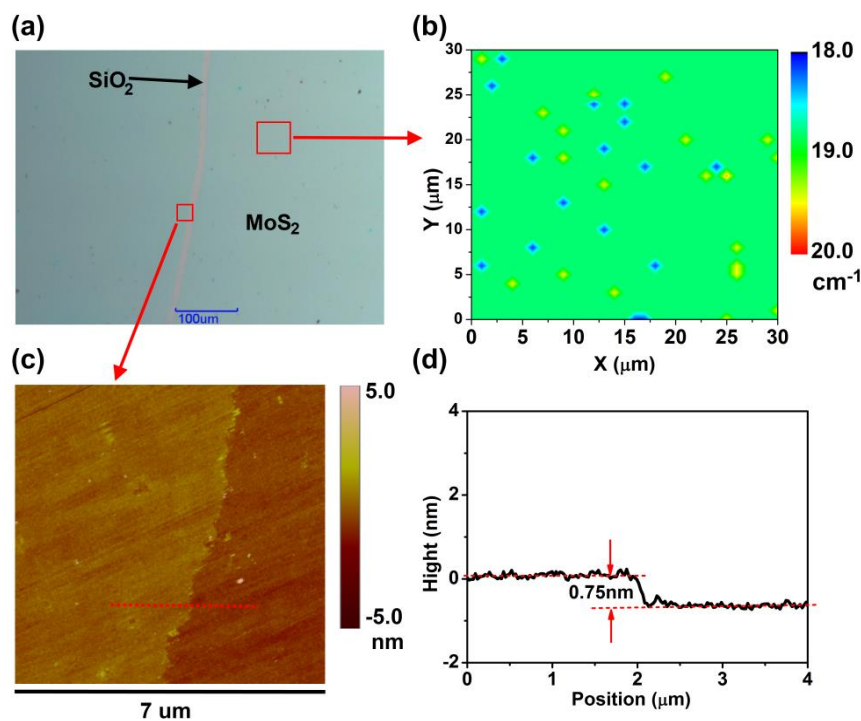

**Figure S2. Characterization of monolayer MoS$_2$ films.** (a) Optical microscope image of a monolayer MoS$_2$ film synthesized on a SiO$_2$/Si substrate. (b) Raman mapping of the frequency separation between A$_{1g}$ and E$_{2g}^1$ peaks. The average value of A$_{1g}$-E$_{2g}^1$ is ~19.0 cm$^{-1}$, indicating that the film is a MoS$_2$ monolayer. (c) Height profile obtained from atomic-force microscopy at the edge of the film. (d) Cross-sectional profile along the red dotted line in (c). The thickness is ~0.75 nm, proving that it is a monolayer MoS$_2$ film.

## Characterization of Charge Carriers.

First, we estimate the carrier concentrations and resistances in the graphene and graphene/$MoS_2$ sheets on $SiO_2$ by Hall-effect measurements. Three samples of graphene on $SiO_2$ and four samples of the graphene/$MoS_2$ heterostructure on $SiO_2$ were measured in dark and ambient air at the room temperature. All of the samples were with the same size (0.5cm square), and each sample was measured four times. As shown in Figure S3(a) and (b), the hole concentrations decreased and the resistances increased in graphene/$MoS_2$ compared with graphene on $SiO_2$, indicating that graphene on $MoS_2$ is less *p*-doped. Second, we prepared the graphene and graphene/$MoS_2$ transistors on a $SiO_2$/Si substrate with the same fabrication processes. The electrical transfer curves for the graphene and graphene/$MoS_2$ transistors are presented in Figure S3(c). The charge neutral point of the graphene/$MoS_2$ transistor shifts to the left, which means that the graphene/$MoS_2$ transistor is less hole-doped, consistent with the results of Hall-effect measurements.

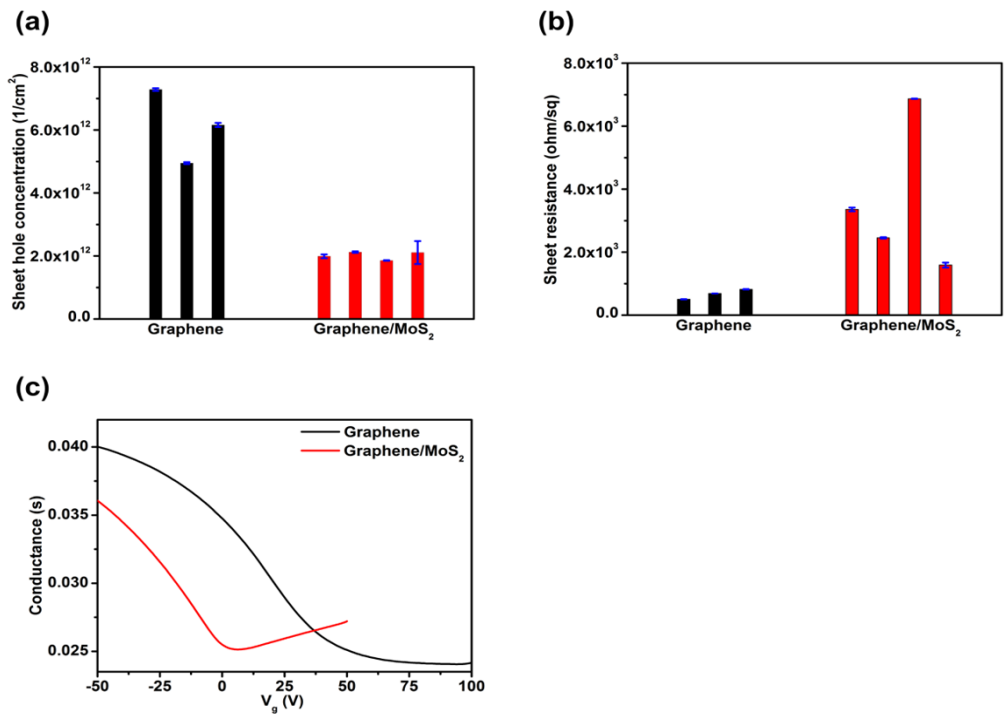

**Figure S3. Characterization of charge carriers.** (a) Hole concentrations and (b) resistances for the graphene sheets on $SiO_2$ and on $MoS_2$/$SiO_2$, respectively. (c) The electrical transfer curves for the graphene and graphene/$MoS_2$ transistors.

## MoS₂ on graphene.

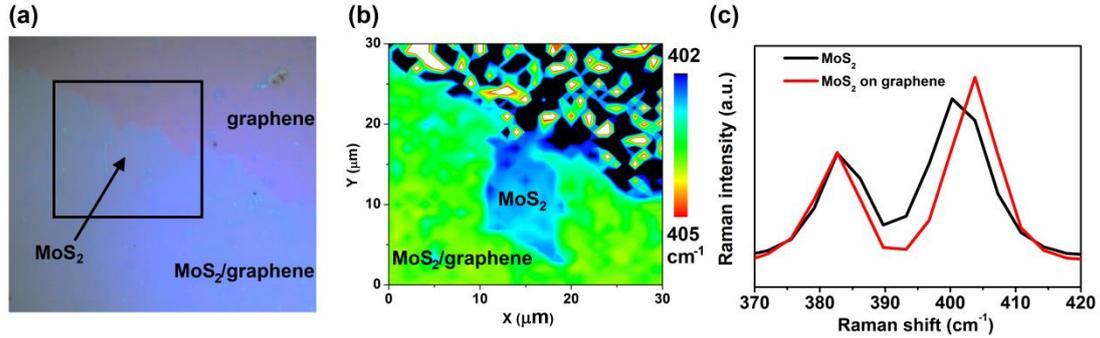

**Figure S4. MoS₂ on graphene stack film.** (a) OM of the heterostructure of MoS₂ on graphene on SiO₂ substrate. (b) The $A_{1g}$ peak position mapping for the area indicated as the black square in (a). (c) The $A_{1g}$ and $E^1_{2g}$ Raman profiles of MoS₂. The right shift of the MoS₂ $E^1_{2g}$ peak after stacked on graphene demonstrates that electrons move to graphene.

## Dynamic photocurrent measurement.

To further exclude the thermal effect, we measured the time-resolved photocurrent driven by different drain voltages at a low laser power density, where the low power density ensures that the fast current self-heating effect becomes more prominent for the thermal desorption process. Figures S5(a) and S5(b) show the normalized photocurrent-time profiles for the graphene/MoS₂ transistor in ambient air and in vacuum. It is observed that the current-time curve obtained at $V_{ds}$ = 0.002 V overlaps with that at $V_{ds}$ = 0.01 V. The photocurrent becomes smaller when $V_{ds}$ is set at 1V (current $I_{ds}$ is ~2×10⁻²A ). If the thermal desorption of dopants from the film is a dominant process, a higher $V_{ds}$ should lead to a larger photocurrent. These results further confirm that thermal desorption is not a dominant process.

To clarify the dynamic mechanism of the photocurrent, we measured the laser power dependence of the time-resolved photoresponse in air and vacuum. The current-time profiles upon switching on the light can be fitted with a double exponential function

$$I = A_1 e^{-\frac{t}{\tau_1}} + A_2 e^{-\frac{t}{\tau_2}} + B$$

where $\tau_1$ and $\tau_2$ are time constants reflecting two different relaxation mechanisms, the values of $A_1$ and $A_2$, represent weighting factors that quantify the relative contributions of each mechanism to the process, and $B$ corrects the baseline. The condition $\tau_1 < \tau_2$ can be assigned so that the first term corresponds to the faster process. As shown in Figure S5(f), the time constant $\tau_1$ almost keeps a constant between 1.8s and 2.5s in both air and vacuum, suggesting that the fast process is intrinsic for the graphene/$MoS_2$ phototransistors. However, the $\tau_2$ in ambient air was smaller than that in vacuum, indicating that this process could be related to the adsorbates.

Using the double exponential function, we fitted the time-resolved photocurrent of another higher mobility sample (Figure S8) and got $\tau_1$~2.5s and $\tau_2$~12s. The fast time constant $\tau_1$~2.5s further confirmed that this process was intrinsic for the graphene/$MoS_2$ phototransistors and could be attributed to the band-to-band excitation. The slow time constant $\tau_2$~12s for the high-mobility sample was smaller than that of the low-mobility sample (Figure S5(f)), indicating that this process is related to disorder and could be attributed to the excitation between the defect or charge impurity states and the band edge.

Recently, we have studied the time-resolved photoresponse for the monolayer $MoS_2$ phototransistor (the results will be published in Advanced Materials) and observed that the adsorbates on the surface of $MoS_2$ in ambient air assisted the recombination of the photogenerated carriers, resulting in a relaxation time of the photocurrent in ambient air smaller than that in vacuum for the monolayer $MoS_2$ phototransistor. So the smaller $\tau_2$ in ambient air for the present graphene/$MoS_2$ phototransistor is likely to result from the adsorbate-assisted recombination of carriers photogenerated between the defect or charge impurity states and band edge.

Figure S4c shows that the time constants $\tau_1$ and $\tau_2$ for the photocurrent do not vary with the drain voltages for both the cases in air and in vacuum, corroborating that thermal desorption is not a dominant process.

In summary, both the laser-wavelength dependence of the photocurrent and the time-resolved photoresponse of different drain currents clearly show that the photocurrent of the graphene/$MoS_2$ phototransistor is not from the thermal effects. Our data suggests two processes for the photocurrent: a fast one from the intrinsic band-to-band excitation and a slow one from the excitation between the disorder states and the band edge.

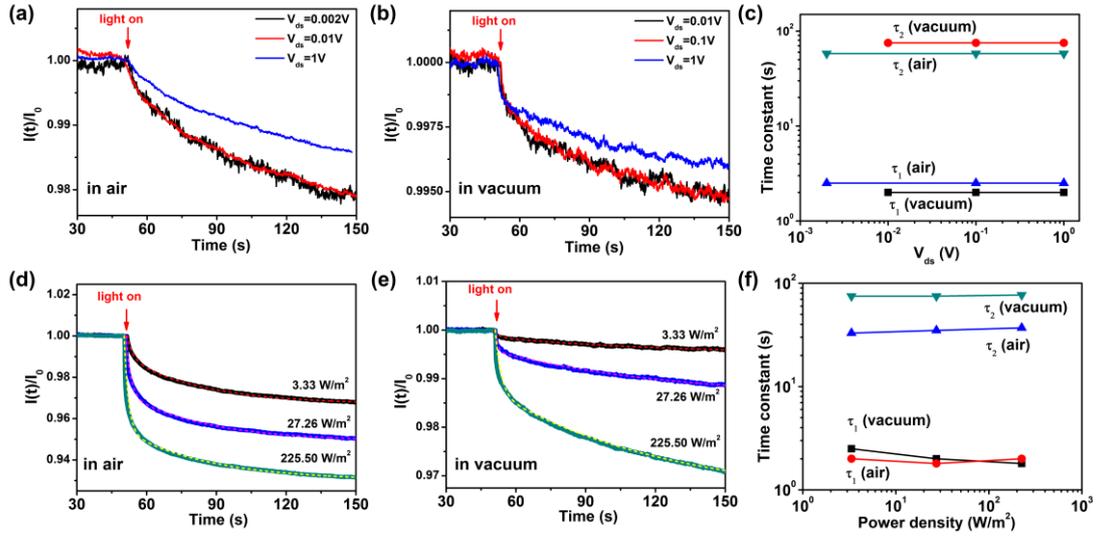

**Figure S5**. **The normalized photocurrent-versus-time profiles** for the graphene/MoS$_2$ transistors measured with various drain voltages in (a) ambient air at a power density of 0.34 W/m$^2$ and (b) in vacuum at a power density of 3.33 W/m$^2$. (c) Time constants $\tau_1$ and $\tau_2$ extracted from figures (a) and (b). Time-resolved photocurrents measured under the exposure of various light intensities (d) in air and (e) in vacuum, respectively, at $V_{ds}$=1V, $V_g$=0V. Dotted lines are the fitted curves by a double-exponential function. (f) Photocurrent time constants $\tau_1$ and $\tau_2$ extracted from figures (d) and (e).

## Photoresponses from CVD graphene.

We have prepared eight CVD graphene transistors fabricated using the same processes. Even with the same substrate and measurement conditions, devices showed some variations in photocurrent behaviour. Six of the devices exhibited no photoresponse at various gate voltages with 532-nm laser illumination at a power density of ~35.21W/m$^2$ as shown in Figure S6. The other two devices showed weak photoresponses as demonstrated in Figure 4d of the text, where the photocurrent was positive when the applied gate voltage was near the Dirac point and negative when the gate voltage was far away the Dirac point (i.e. $V_g$< -58V ). The reason for the device variations is still unknown at this moment, and it would require more investigation.

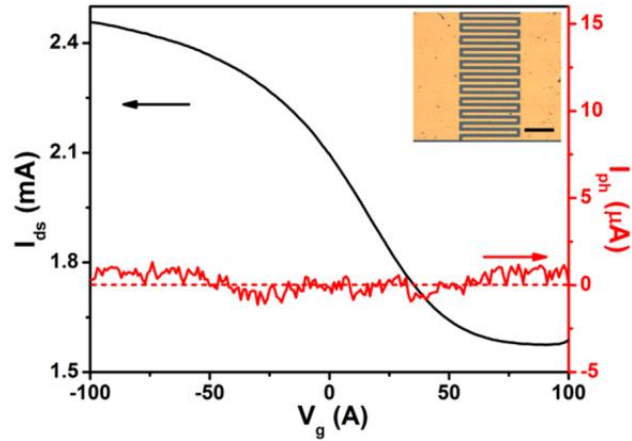

**Figure S6. Photocurrent ($I_{ph}=I_{light}-I_{dark}$) of graphene transistors**. Photocurrent (red) and dark source-drain current (black) as a function of gate voltage. Inset: The optical micrograph of the device, and the black scale bar is 100μm. A 532nm laser with a power density 35.21W/m$^2$ was used. ($V_{ds}$=0.1V)

## Contact between the graphene and MoS$_2$ monolayers.

After the monolayer graphene and monolayer MoS$_2$ contact with each other, the electrons are injected into graphene based on our experimental results. Thus, the conventional theory of ideal metal-semiconductor contacts predicts an energy-band bending at the interface as shown Figure S7, and the direction of the built-in electric field would be from the MoS$_2$ to graphene. When illumination on the graphene/MoS$_2$ transistors, the photo-generated electrons should flow into MoS$_2$ along the bending energy band at the interface. However, this is opposite to our experimental results.

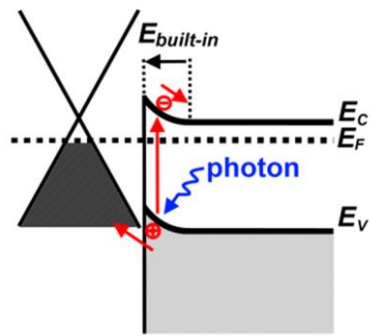

**Figure S7. Contact between graphene and MoS$_2$ monolayers**. The energy band diagram was plotted according to the conventional theory of ideal metal-semiconductor contacts.

## Photocurrent decay in graphene/MoS₂ stack films.

The *I*-t curves obtained after the light is switched off can be well fitted to a double exponential function:

$$I = A_1(1 - e^{-\frac{t}{\tau_1}}) + A_2(1 - e^{-\frac{t}{\tau_2}}) + B$$

where $\tau_1$ and $\tau_2$ are time constants reflecting two different relaxation mechanisms, the values of $A_1$ and $A_2$, represent weighting factors that quantify the relative contribution of each mechanism to the process, and *B* corrects the baseline. The condition $\tau_1 < \tau_2$ can be assigned so that the first term corresponds to the faster process. We obtain $\tau_1 \sim$ 12 sec and $\tau_2 \sim$ 1350 sec for the curve, where the fast process is ascribed to the band-to-band recombination of the photoexcited electron-hole pairs and the slower one is related to the recombination through the trap centers (defects or charge impurities), indicating the presence of charge impurities associated with the MoS₂ layers. Note that the charge impurities may be related to the presence of perpendicular external field discussed in Figure 6 (in the text).

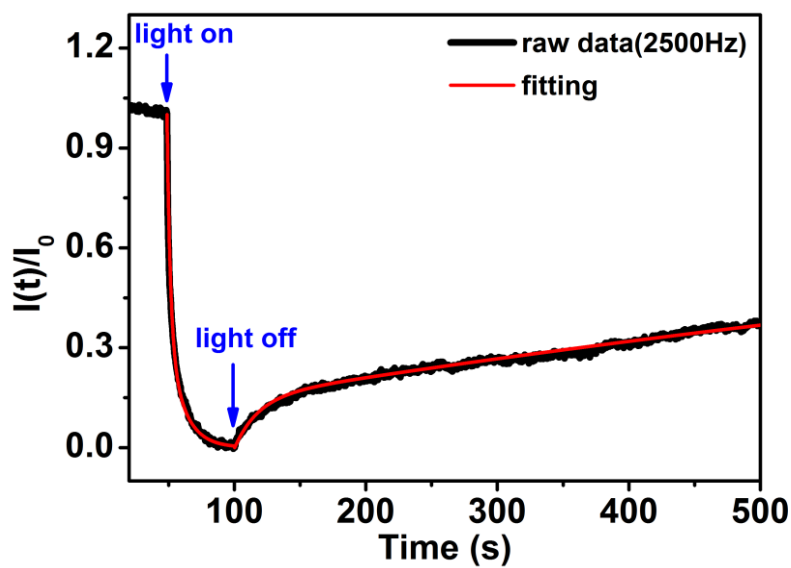

**Figure S8**. **The time-resolved photocurrent for the graphene/MoS₂ transistor** in ambient air with a chopper frequency of 2500Hz. The excitation wavelength is 650 nm and power density is 1.74W/m² at $V_g = 0$V and $V_{ds} = 1$V.